\documentclass[preprint, 3p, times, sort&compress]{elsarticle}

\usepackage{algpseudocode}
\usepackage{algorithm}
\usepackage{comment}
\usepackage{amsmath,amssymb}
\usepackage{amsmath,amssymb}
\usepackage{type1cm}
\usepackage{mathrsfs}
\usepackage{url}
\usepackage{graphicx}
\usepackage{color}
\usepackage{comment}

\newcommand\HL[1]{{\color{black}#1}}
\usepackage[columnwise]{lineno}

\usepackage{listings}

\journal{Information Sciences}
\begin{document}

\begin{frontmatter}



\title{
Auto-generation of a centerline graph from a geometrically complex roadmap of real-world traffic systems using a hierarchical quadtree for cellular automata simulations
}



\author[RCAST]{Satori Tsuzuki}
\ead{tsuzuki.satori@mail.u-tokyo.ac.jp}

\author[RCAST]{Daichi Yanagisawa}
\ead{tDaichi@mail.ecc.u-tokyo.ac.jp}

\author[RCAST]{Katsuhiro Nishinari}
\ead{tknishi@mail.ecc.u-tokyo.ac.jp}

\address[RCAST]{Research Center for Advanced Science and Technology, The University of Tokyo, 4-6-1, Komaba, Meguro-ku, Tokyo 153-8904, Japan}

\begin{abstract}
This paper proposes a method of auto-generation of a centerline graph from a geometrically complex roadmap of real-world traffic systems by using a hierarchical quadtree for cellular automata simulations. Our method is summarized as follows. First, we store the binary values of the monochrome image of target roadmap (one and zero represent the road and the other areas, respectively) in the two-dimensional square map. Second, we recursively divide the square map into sub-leafs by a quadtree until the summed-up value of pixels included inside the leaf becomes equal to or less than one. Third, we gradually remove the distal leaves that are adjacent to the leaves whose depths are shallower than the distal leaf. After that, we trace the remaining distal leaves of the tree using Morton's space-filling curve, while selecting the leaves that keep a certain distance among the previously selected leaves as the nodes of the graph. Finally, each selected node searches the neighboring nodes and stores them as the edges of the graph. We demonstrate our method by generating a centerline graph from a complex roadmap of a real-world airport and by carrying out a typical network analysis using Dijkstra's method.
\end{abstract}


\end{frontmatter}





\section{INTRODUCTION}
\begin{figure}[t]
\vspace{-1.0cm}
\hspace{+1.5cm}
\includegraphics[width=\textwidth, clip, bb= 0 0 1280 720]{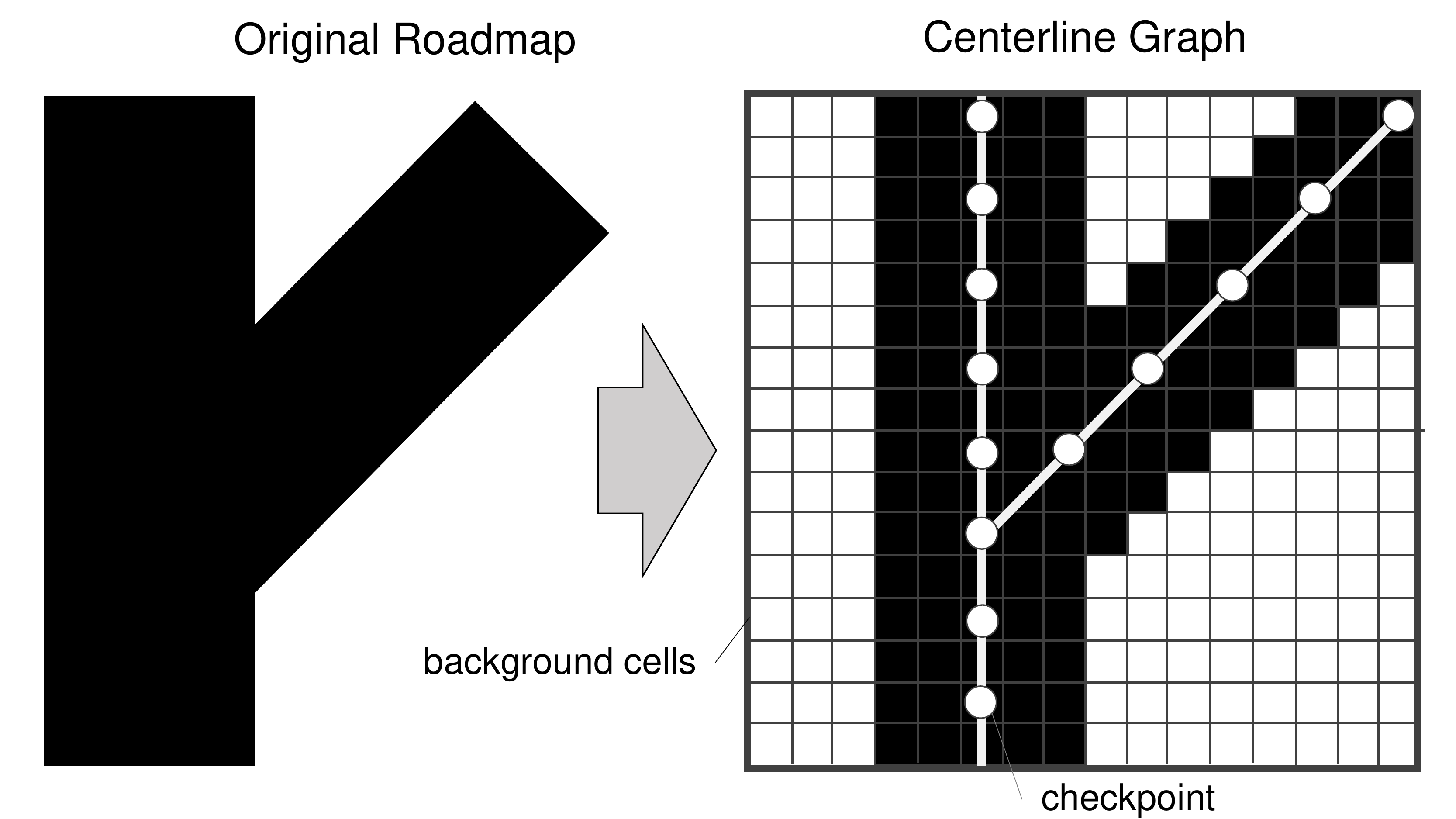}
\caption{A schematic of the generation of a centerline graph from a roadmap for traffic simulations using CA.}
\label{fig:schemclgraph}
\end{figure}

Cellular automaton (CA), which was established by Neumann~\cite{Neumann:1966:TSA:1102024}, has attracted scientists and researchers for years. Notably, after Wolfram found the Elementary CA Rule 184~\cite{RevModPhys.55.601}, the various families of CA have been acknowledged in the field of traffic flow problems. In recent vehicular traffic simulations, cellular automata have taken advantage of fine-grained background cells to represent the complex geometries of real-world systems. As depicted in Fig.\ref{fig:schemclgraph}, the system selects representative cells from the background cells as the checkpoints and constructs a graph by connecting them. Because the vehicles move on the centerlines of the routes of the graph during simulations~\cite{YAMAMOTO2007654, CDA14, 1901.10390, 6248222, FELICIANI2016135}, it is necessary to extract a centerline graph from a roadmap for pre-processing in realistic vehicular traffic simulations.

\begin{figure}[t]
\vspace{-1.0cm}
\hspace{+1.5cm}
\includegraphics[width=\textwidth, clip, bb= 0 0 1280 720]{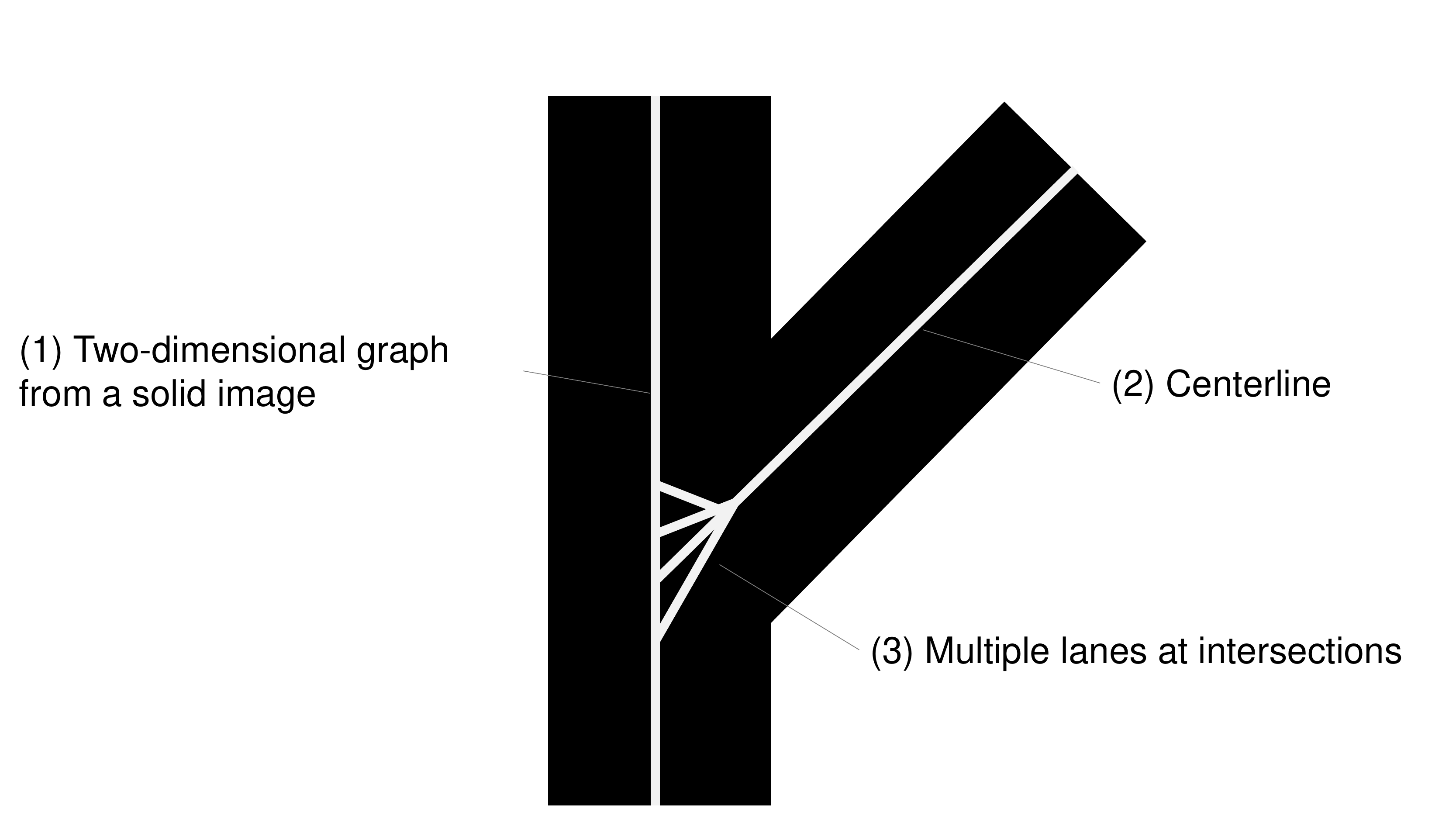}
\caption{A schematic of the characteristics of the target graph.}
\label{fig:diffprevandthis}
\end{figure}

\HL{This study aims to conduct a preliminary research as the first step toward the auto-generation of the network graph used in practical traffic CA simulations. Promising approaches for centerline extraction from twisted paths can be found in computed tomography~\cite{1176632, 8283818}, environmental and urban engineering~\cite{Haunert2008, Yuqing09theshortest, rs6099014, doi:10.1559/152304000783547966, ZHONG201850, doi:10.1080/13658816.2014.915401}, distance-measurement based learning algorithms\cite{hong2015multimodal, wang2018application, mao2019multiobjective, noormohammadi2019multi, han2019efficient, yu2019spatial, yu2014semantic, yu2016deep}, and topological thinning~\cite{CHU1986207, RONSE198631, MA1994328, 22716b0af8f74680a3cd485e2932e1cf, Passat:2010:TPT:1745807.1745813, doi:10.1002/ima.20272, PALAGYI2006974, Abeysinghe2009, WONG20061379}. Here, focusing our interest on the problems of vehicular traffic simulations restricts the geographic shape of the target graphs, which are summarized as follows: (1) A two-dimensional graph is sufficient. (2) The graph passes around the centerline of the original road map when it is on straight roads. (3) The graph has multiple lanes at intersections in order to represent various movements of vehicles around corners in realistic cases. Figure~\ref{fig:diffprevandthis} illustrates the characteristics of (1) to (3) for reference. By considering these restrictions in this study, we can develop new techniques that comprises both straightforwardness and computational efficiency compared to the aforementioned related studies.}

\HL{In this paper, we propose a method to generate a centerline network graph from geometrically complex roadmaps of real-world systems by dynamically changing the structure of a hierarchical tree, by taking advantage of the characteristics of solid images.} Our proposed method can be summarized as follows. First, we store the binary values of the monochrome image of target roadmap (one and zero represent the road and the other areas, respectively) in the two-dimensional square map. Second, we recursively divide the square map into sub-leafs by a quadtree until the summed-up value of pixels included inside the leaf becomes equal to or less than one. Third, we gradually remove the distal leaves that are adjacent to the leaves whose depths are shallower than that of the distal leaf, during the shrinking process. After that, we trace the remaining distal leaves of the tree using Morton's space-filling curve~\cite{konrad:inria-00331382, SFCDecomp1997-Aluru, Tsuzuki:2016:EDL:3019094.3019095, 4658143}, while selecting the leaves that keep a certain distance from among the previously selected leaves as the nodes of the graph. Finally, each selected node searches the neighboring nodes and stores them as the edges of the graph.
In short, we attempt to generate a centerline graph just by releasing the distal leaves. 

\HL{
Our method significantly improves computational efficiency compared to the existing algorithms. For example, a grid search technique associated with the distance functions requires a computational time of $O(N^{\rm 2})$, where $N$ indicates the number of pixels in one direction of the square map. If a graph generation algorithm includes the traveling salesman problem (TSP), it requires a computational time of $O(N!)$ to get the exact solutions, and $O(N^{\rm 2} {\rm 2}^{N})$ to get approximations using a dynamic programming algorithm in typical cases\cite{Hoffman2013}, where $N$ indicates the number of nodes. Meanwhile, the topological thinning-based algorithm reported in~\cite{WONG20061379}, which is the closest to our approach, requires a computational time of $O(N^{\rm 2})$, where $N$ indicates the number of leaves. The reason for the high computational costs is because the algorithm in~\cite{WONG20061379} carries out the reconstruction of a quadtree in every iterative step of the shrinking process. In contrast, our method requires a one-time recursive tree construction. Substantially, our algorithm requires a computational time of $O(k {\rm log} N)$ to $O(k N)$ for a shrinking process, where $k$ indicates a scalar constant value. In this manner, our method not only has significant straightforwardness but also shows computational efficiency equal to that of the fastest state-of-the-art algorithms such as the one in \cite{yu2014semantic}. 
Note that, based on these advantages, we determine the value corresponding to the parameter $k$ experimentally, as discussed in Section~2.4.}

The remainder of this paper is structured as follows. Section 2, which is divided into four parts, describes our method. In Section 3, we analyze the shortest path search using Dijkstra's algorithm. Section 4 summarizes our results and concludes this paper.

\section{METHOD} 
\begin{figure}[t]
\vspace{-4.0cm}
\hspace{-1.7cm}
\includegraphics[width=1.62\textwidth, clip, bb= 0 0 1280 720]{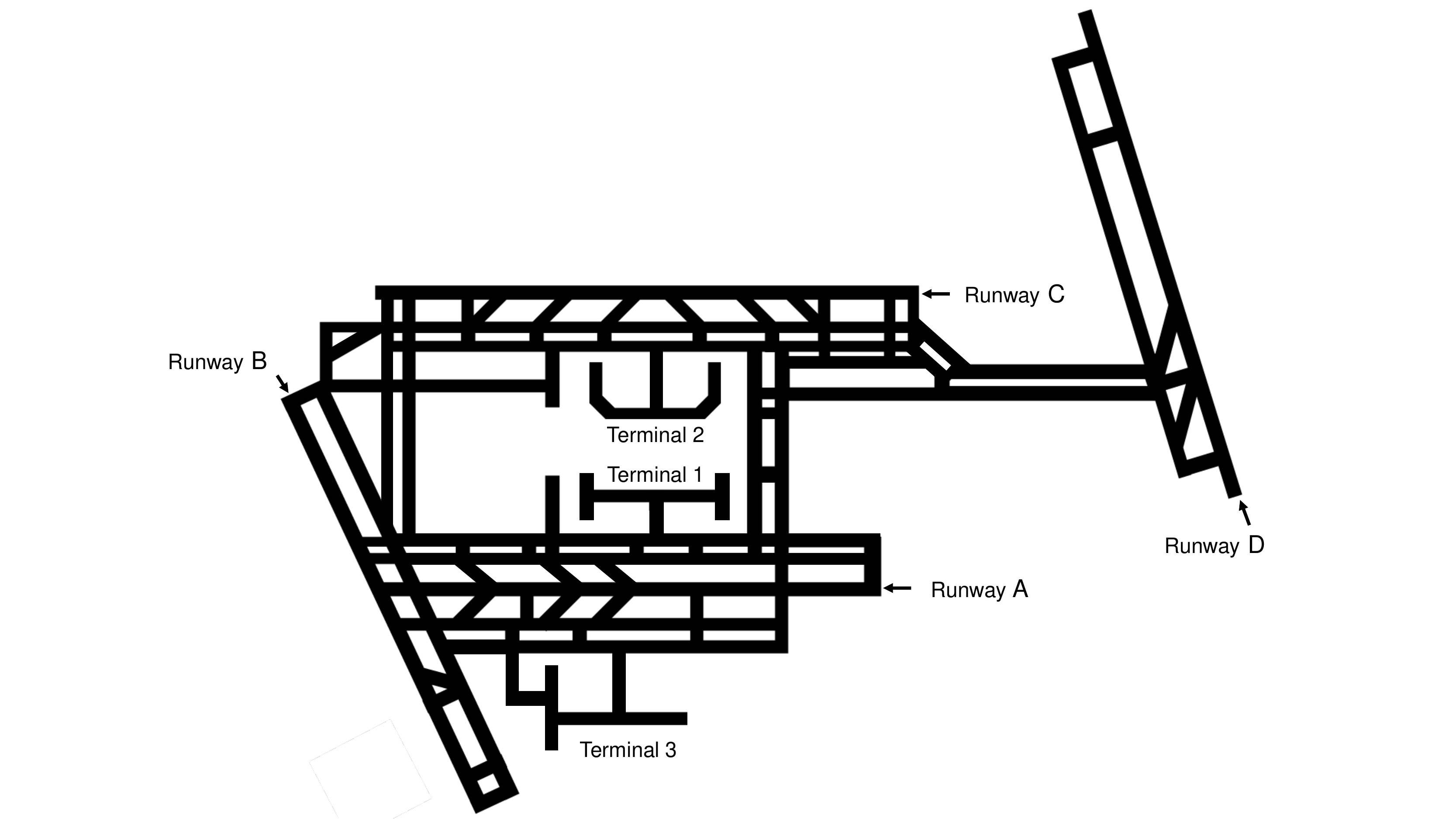}
\caption{Targe roadmap abstracted from a real-world airport, Tokyo International Airport in Japan.}
\label{fig:TargetProblem}
\end{figure}

\begin{figure}[t]
\vspace{-1.5cm}
\hspace{+2.0cm}
\includegraphics[width=\textwidth, clip, bb= 0 0 1280 1280]{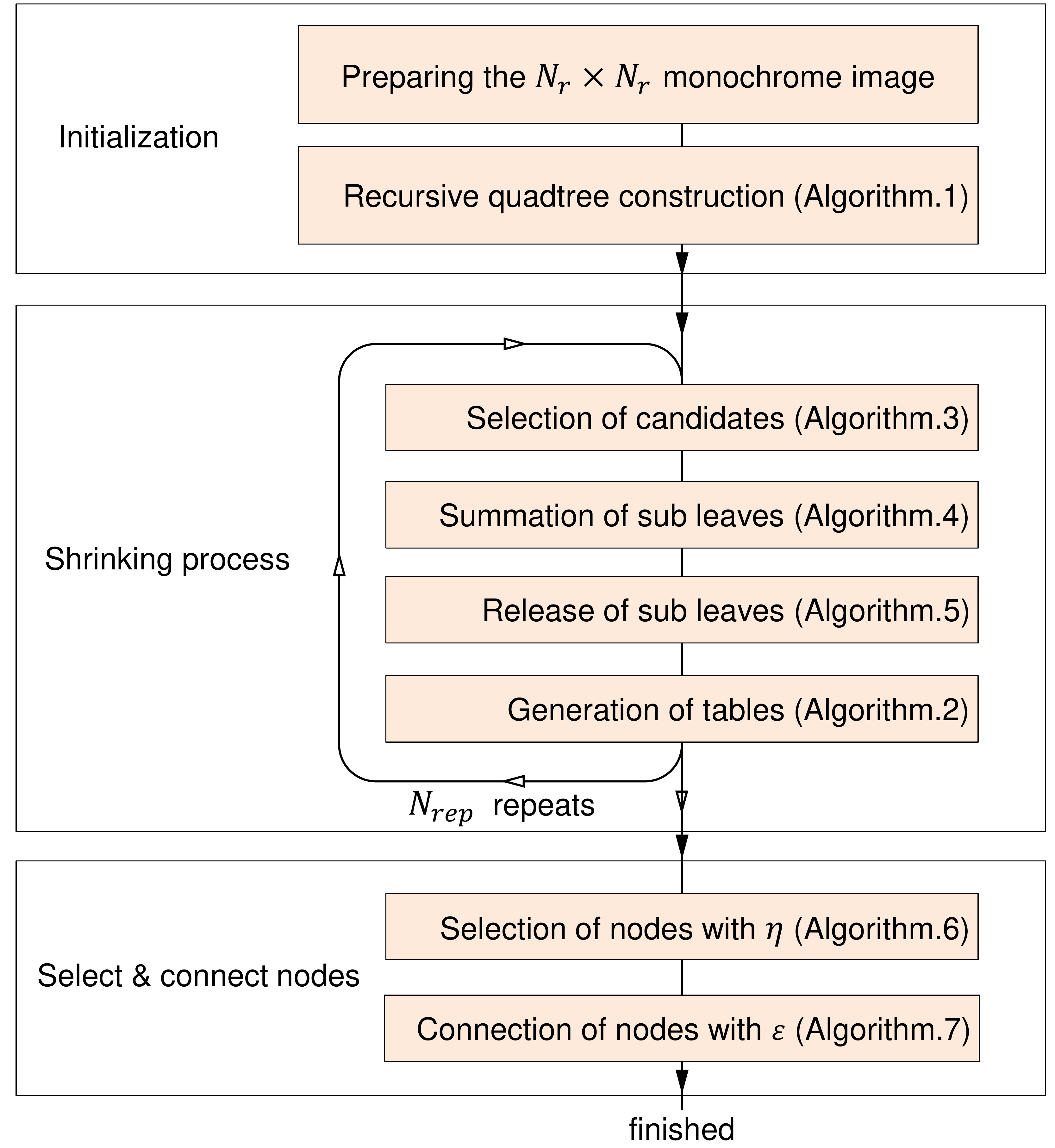}
\caption{A flowchart of our method.}
\label{fig:flowchart}
\end{figure}

Figure~\ref{fig:TargetProblem} shows the target roadmap, which we abstracted from a real-world airport, namely the Tokyo International Airport in Japan. We can divide our method into four processes: the construction process of a recursive tree on the roadmap, the release of distal leaves, the selection of nodes, and the connection of nodes by tracing the distal leaves using a space-filling curve. 
\HL{Figure~\ref{fig:flowchart} shows a flowchart of our proposed method.}

\subsection{Data structure of leaves}
We store each element of the monochrome image of the roadmap depicted in Fig.~\ref{fig:TargetProblem} in a corresponding cell of the two-dimensional $N_{\rm r}\times N_{\rm r}$ array. Here, the value of one or zero in each cell represents the road and the other areas, respectively. The parameter $N_{\rm r}$ denotes the number of cells in either of the vertical or horizontal directions in the original roadmap, whichever is larger. After that, we recursively divide the square map into $2\times2$ sub-leafs by a quadtree, each leaf of which has the following data structure:
\begin{lstlisting}[mathescape=true, numbers=none, label=leafstr, caption= An example of the leaf structure.]
	Class Leaf {
		Vector   $L_{\rm min}$; // Minimum coordinates. 
		Vector   $L_{\rm max}$; // Maximum coordinates. 
		int    $D$; // Depth of the leaf.   
		int    $R$; // Ordering by parent leaf.
		bool   $F_{\rm end}$; // Distal leaf or not.
		bool   $F_{\rm del}$; // To delete (release) or not.
		int    $S_{\rm del}$; // Sum of $\color[rgb]{0,0.5,0}{F_{\rm del}}$.
		bool   $F_{\rm chk}$; // Checkpoint or not.
		Leaf** child; // Pointer to child leaves.
		Leaf** parent; // Pointer to parent leaf.
	}
\end{lstlisting}
Here, the vectors $L_{\rm min}$ and $L_{\rm max}$ indicate the minimum and maximum coordinates of the leaf and correspond to $(0,0)$ and $(N_{\rm r}, N_{\rm r})$ at a depth of zero, respectively. The parameter $D$ represents the depth of the leaf. The parameter $R$ indicates the ordering index designated by their parent leaf, which determines the geometric location of the leaf among the bundle of sub-leafs. 
A space-filling curve is a recursive curve that fills a space or area with a single stroke of a brush~\cite{konrad:inria-00331382, SFCDecomp1997-Aluru}; by making the ordering indices of sub-leafs correspond to the space-filling curve, we can trace all the distal leaves using a sequential single curve. Here, the ordering pattern $R$ is individually determined by the type of space-filling curve. There exist several types of space-filling curves\cite{bader2012space}. In this paper, we use Morton's curve (Z-curve) because of its simplicity. The Morton's curve gives indices between $0$ and $4$ to the sub-leafs; therefore, the trajectory of traversing the sub-leafs produces a ``Z'' shape. Hence, the parameter $R$ is always a number from $0$ to $4$. 

The binary parameter $F_{\rm end}$ represents whether the leaf is a distal leaf or not. We determine the $F_{\rm end}$ of each leaf in Algorithm~\ref{pesuedo-constTree}, which is shown later. Meanwhile, we use the binary parameter $F_{\rm del}$ and parameter $S_{\rm del}$ to judge whether we can release the distal leaf or not in Algorithm~5. In addition, the parameter $F_{\rm chk}$ indicates whether we can select the leaf as a node of the graph or not, and we obtain $F_{\rm chk}$ of each leaf using Algorithm~6.

\subsection{Recursive quadtree construction}
\begin{figure}[t]
\vspace{-2.5cm}
\hspace{+3.5cm}
\includegraphics[width=0.75\textwidth, clip, bb= 0 0 960 960]{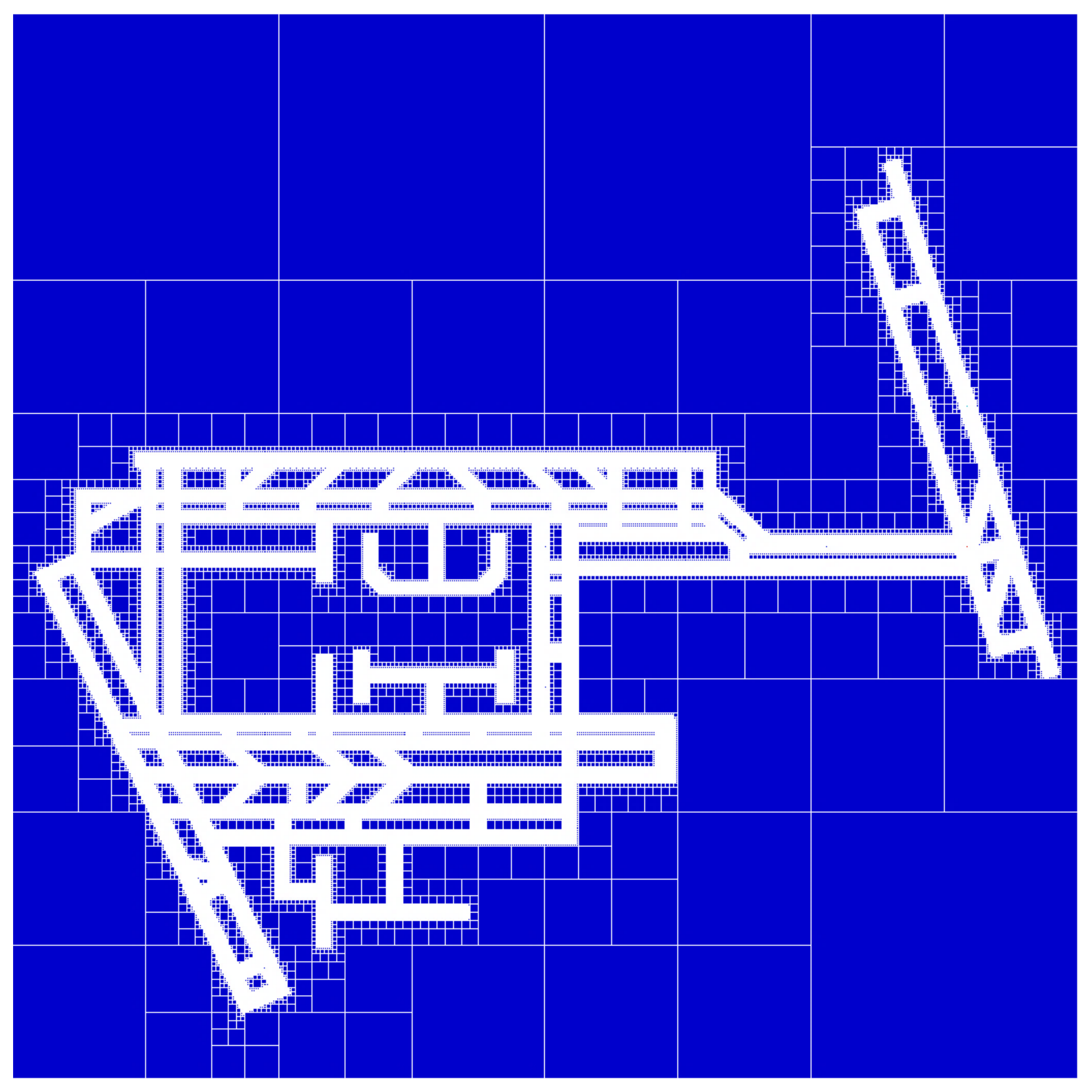}
\caption{A hierarchically structured mesh constructed for the roadmap in Fig.\ref{fig:TargetProblem}.}
\label{fig:QuadTreeMesh}
\end{figure}

Algorithm~\ref{pesuedo-constTree} describes the procedure of the recursive quadtree construction. Lines2-9 show the process of initialization of all the parameters of each leaf. In the process after line9, we recursively divide the $N_{\rm r} \times N_{\rm r}$ map into $2 \times 2$ sub-leafs by a quadtree. The parameter $B_{\rm p}$ indicates the state of a cell, which becomes one when the leaf is inside the road and zero in other cases. In line10, we define $B_{\rm p}^{\rm s}$ as the summed up value of $B_{\rm p}$ inside the leaf. We keep performing recursive division of the tree until the value of $B_{\rm p}^{\rm s}$ in each leaf becomes equal to or less than one. As a result, each of the deepest distal leaves corresponds to a single cell on the road.
Figure~\ref{fig:QuadTreeMesh} shows a hierarchically structured mesh constructed on the roadmap of the airport depicted in Fig.~\ref{fig:TargetProblem}. Here, two important observations are made. First, the distal leaves covering the cells on the inner edges of the roads always have the maximum depth of the tree. Second, each cell on the inner edges of the roads are always adjacent to the other leaves that have a shallower depth than the cell. By utilizing these characteristics, we retain only the leaves located around the centerline by releasing leaves through the processes in Algorithms 3 to 5. 

\begin{figure}[t]
\vspace{-2.5cm}
\hspace{+3.5cm}
\includegraphics[width=0.75\textwidth, clip, bb= 0 0 960 960]{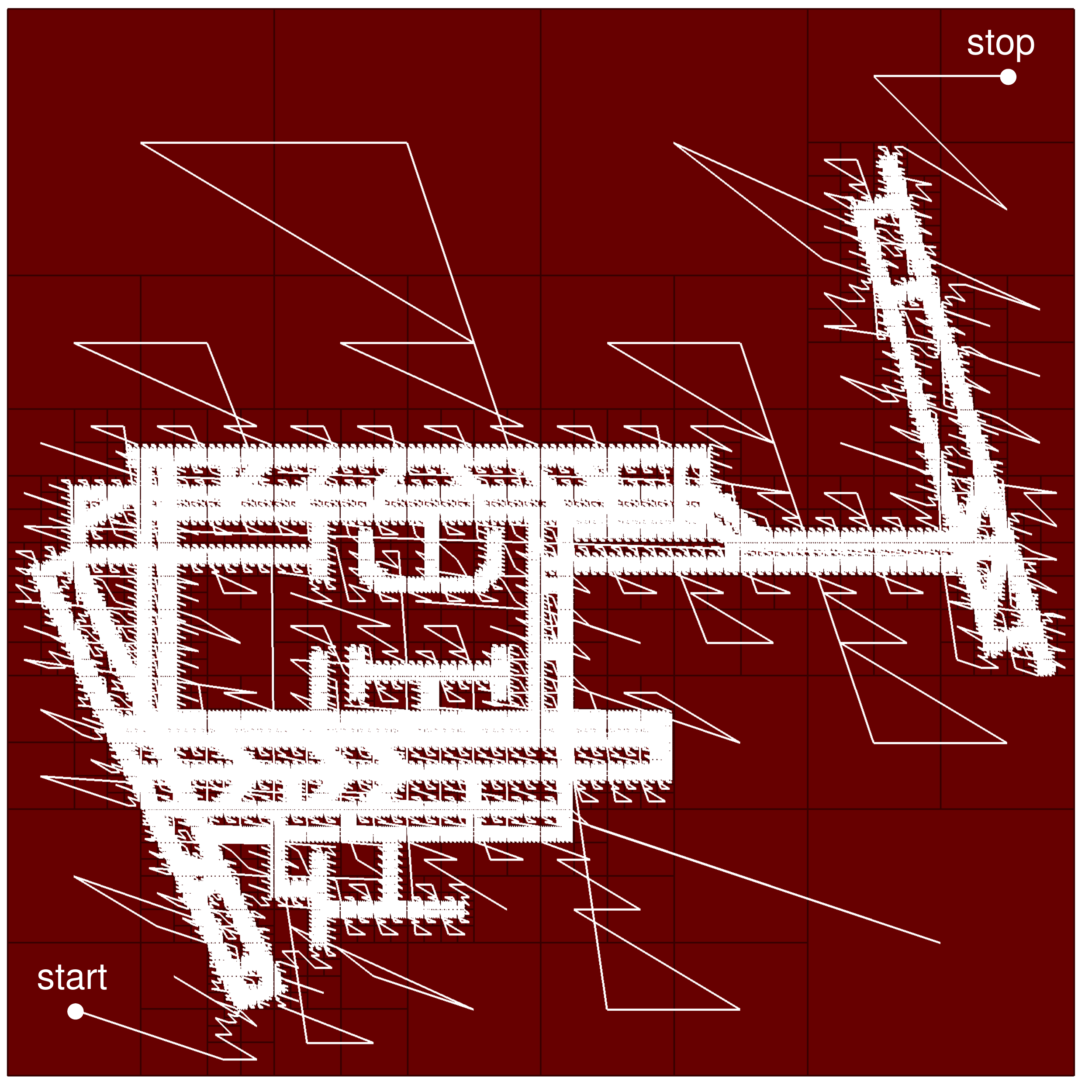}
\caption{Tracing of all the distal leafs by using Morton's space-filling curve.}
\label{fig:MortonCurve}
\end{figure}

Figure~\ref{fig:MortonCurve} shows the process of tracing all the distal leaves by ``a single stroke of a brush'' using Morton's curve. Here, for easy access to the distal leaf of the tree, we introduce a conventional technique of a pointer table as shown in Algorithm~\ref{pesuedo-pointertable}. We can set a sequential number for all the distal leaves by incrementing the number every time Morton's curve traverses two different distal leaves. We can create one more $N_{\rm r} \times N_{\rm r}$ array and fill the area paved by a distal leaf with the index of the leaf. Similarly, it is possible to store the pointer of each distal leaf in an $N_{\rm r}\times N_{\rm r}$ array, as shown in line8. This pointer table not only makes it easier to access all the distal leaves but also simplifies all the algorithms. Additionally, we store a pointer of the parent leaf in line 2, which we use in the process of releasing the leaves in Algorithm~5.

\begin{algorithm}[t]
\caption{Recursive Tree Construction}\label{pesuedo-constTree}
\begin{algorithmic}[1]
\Function{Tree}{{$L_{\rm min}$, $L_{\rm max}$, $D$, $R$}}
\State ${\rm Leaf.}L_{\rm min}$~$\leftarrow$~$L_{\rm min}$ 
\State ${\rm Leaf.}L_{\rm max}$~$\leftarrow$~$L_{\rm max}$ 
\State ${\rm Leaf.}D$~$\leftarrow$~$D$ 
\State ${\rm Leaf.}R$~$\leftarrow$~$R$ 
\State ${\rm Leaf.}F_{\rm end}$~$\leftarrow$~false 
\State ${\rm Leaf.}F_{\rm del}$~$\leftarrow$~false
\State ${\rm Leaf.}S_{\rm del}$~$\leftarrow$~0
\State ${\rm Leaf.}F_{\rm chk}$~$\leftarrow$~false
\State Define $B_{\rm p}^{\rm s}$ as the sum of $~B_{\rm p} \in [L_{\rm min}, L_{\rm max}]$ 
\If{$B_{\rm p}^{\rm s} \le 1$}
\State ${\rm Leaf.}F_{\rm end}$~$\leftarrow$~true 
\For{k~=~0,1...$N_{\rm child}-1$}
\State $\rm Leaf.child$[k]~$\leftarrow$~NULL 
\EndFor
\Else
\For{k~=~0,1...$N_{\rm child}-1$}
\State Set [$L_{\rm min}^{\rm k}, L_{\rm max}^{\rm k}$] and $R^{\rm k}$~by~$R$
\State $\rm Leaf.child$[k]~$\leftarrow$~\Call{Tree}{{$L_{\rm min}^{\rm k}$, $L_{\rm max}^{\rm k}$, $D+1$, $R^{\rm k}$}}
\EndFor
\EndIf
\State \Return $\rm Leaf$
\EndFunction
\end{algorithmic}
\end{algorithm}

\begin{algorithm}[t]
\caption{Generation of Table}\label{pesuedo-pointertable}
\begin{algorithmic}[1]
\Function{genTable}{$\rm Leaf$, $\rm parent$}
\State ${\rm Leaf.}{\rm parent}$~$\leftarrow$~$\rm parent$ 
\If{${\rm Leaf.}F_{\rm end}$~$=$~${\rm false}$} 
\For{k~=~0,1...$N_{\rm child}-1$}
\State \Call{genTable}{$\rm Leaf.child[k]$, $\rm Leaf$}
\EndFor
\Else
\ForAll{$\rm j$ s.t. ${\rm Leaf.}L_{\rm min}.y\leq {\rm j}\leq {\rm Leaf.}L_{\rm max}.y$}
\ForAll{$\rm i$ s.t. ${\rm Leaf.}L_{\rm min}.x\leq {\rm i}\leq {\rm Leaf.}L_{\rm max}.x$}
\State Table[i][j] $\leftarrow$ Leaf
\EndFor
\EndFor
\EndIf
\EndFunction
\end{algorithmic}
\end{algorithm}

\subsection{Release of distal leaves}
\begin{figure}[t]
\vspace{-1.5cm}
\hspace{+0.8cm}
\includegraphics[width=1.2\textwidth, clip, bb= 0 0 1707 960]{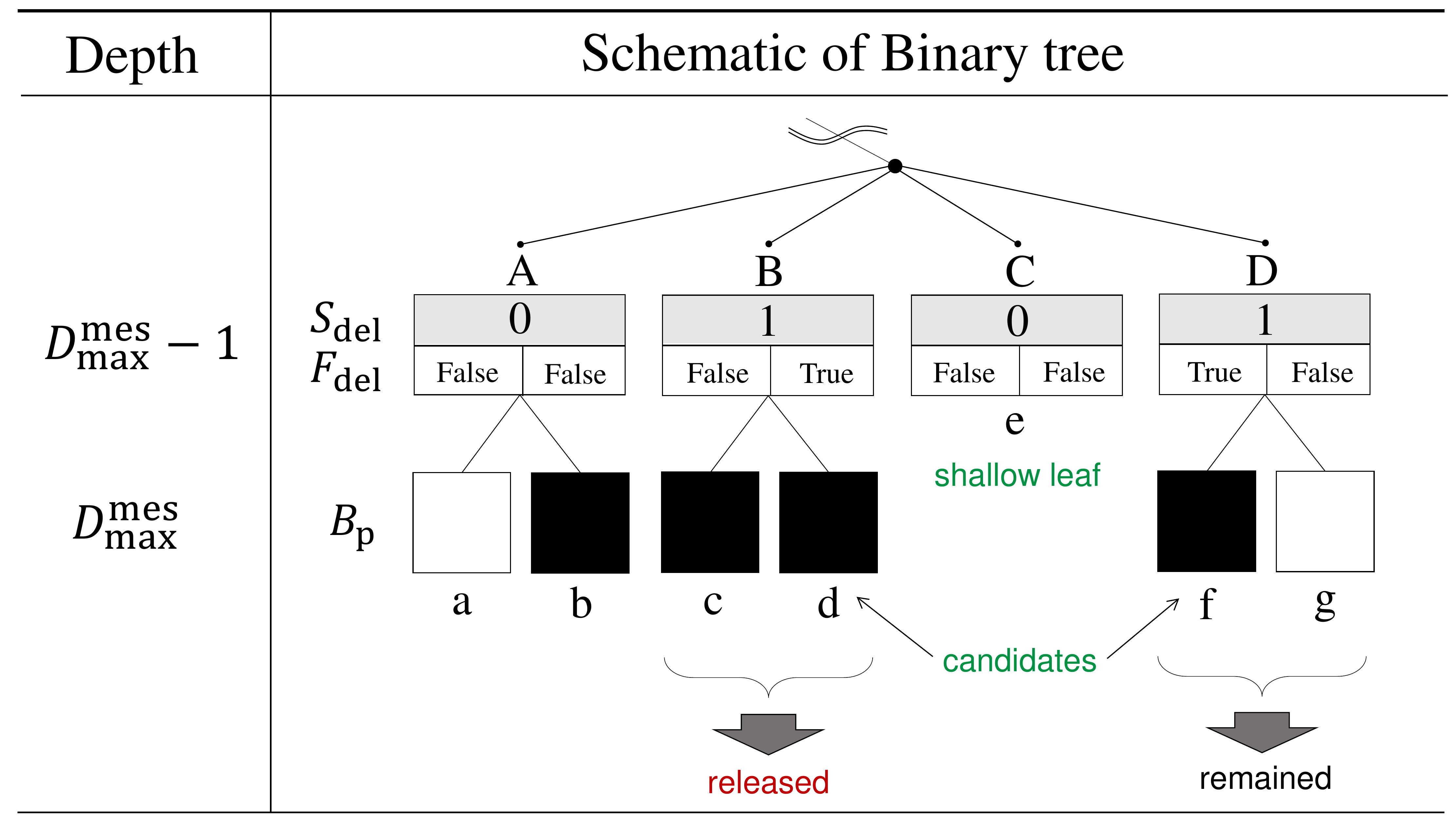}
\caption{A schematic explanation of releasing the distal leaves.}
\label{fig:SchemRelLeaf}
\end{figure}

Figure~\ref{fig:SchemRelLeaf} illustrates a schematic of releasing the distal leaves. Here, we consider a one-dimensional case using a binary tree for easy explanation. The parameter $D_{\rm max}^{\rm mes}$ indicates the maximum depth of the tree measured by tracing all the distal leaves using Morton's curve. First, we find out the leaves that are adjacent to the leaves having a shallower depth by one level compared to the deepest leaf. We flag them as the candidates and change their $F_{\rm del}$ from false to true. In this case,  we flag the leaf $d$ and leaf $f$ as the candidates because leaf $e$ has a shallower depth. We need to prevent the tree from over-releasing leaves because the tree releases them by units of a bundle of sub-leafs. We count the number of $F_{\rm del}$ whose state is true in every bundle of sub-leafs and store it in the $S_{\rm del}$ of their parent leaf. When the $S_{\rm del}$ of a leaf is zero, the leaf has no candidate sub-leaf to be released. Hereinafter, we refer to such a leaf as ``a stable leaf.'' Next, leaf B searches the state of neighboring leaves (A and C). Because leaf B is connected with a stable leaf A, we keep the distal leaf $d$ as a candidate sub-leaf to be released. On the other hand, we have to exclude leaf $f$ from the list of candidates regardless of the state of leaf D because leaf $f$ connects with a vacant leaf $g$, which is located outside the roads.
In the end, leaf B, a bundle comprising leaf $c$ and leaf $d$, is released.

Algorithm~\ref{pesuedo-cand} shows the procedure of selecting the candidates to be released. First, we detect one of the distal leaves by the process in lines 2-6. After finding a distal leaf, we examine the depth of the four neighboring leaves existing in the vertical or horizontal direction of the distal leaf as described in lines12-26. If we find that at least one of the four neighboring leaves has a shallower depth than $D_{\rm max}^{\rm mes}$, we change the $F_{\rm del}$ of the leaf from false to true as shown in lines27-29. After that, in Algorithm~\ref{pesuedo-sumbundleleafs}, we count the number of $F_{\rm del}$ whose state is true in every bundle of sub-leafs and store it in the $S_{\rm del}$ of their parent leaf as described in lines 9-14. As aforementioned, when the value of $S_{\rm del}$ of a leaf becomes zero, the leaf has no candidate sub-leaf to be released (the leaf can be considered a ``stable leaf'').

Finally, we remove leaves according to the procedure in Algorithm~\ref{pesuedo-release}. We detect one of the parents of the distal leaves through the process in lines2-7. After finding a parent of a distal leaf, we descend to each of the four child leaves of the leaf as described in line10. We then examine the value of $S_{\rm del}$ of each parent of the leaves neighboring each child leaf as described in lines13-24. When the value of $S_{\rm del}$ becomes zero, a bundle of the leaves including the neighboring leaf of the child leaf remains; we conclude that the child leaf is connected to ``a stable leaf''. In this case, we change the $F_{\rm nbr}$ from false to true. On the other hand, we check the value of $F_{\rm del}$ of each child leaf. If at least one of $F_{\rm del}$ of child leaves is true, we change $F_{\rm cnd}$ from false to true in lines27-32. As long as both $F_{\rm nbr}$ and $F_{\rm cnd}$ are true, we release all the child leaves as described in lines33-39.

We set the parameter $F_{\rm stop}$ to be false at the root of the tree. By examining the value of $F_{\rm stop}$ after performing Algorithm~\ref{pesuedo-release}, we can judge whether the release of leaves occurs at least once or not. We repeat the procedures in Algorithms~\ref{pesuedo-cand} to \ref{pesuedo-release} to shrink the area around the centerline. It should be noted that it is necessary to update the pointer table ${\rm Table (i,j)}$ and $D_{\rm max}^{\rm mes}$ at every step after executing the processes in Algorithms \ref{pesuedo-cand} to \ref{pesuedo-release}. 

\begin{figure}[t]
\vspace{-4.5cm}
\hspace{+3.0cm}
\includegraphics[width=0.8\textwidth, clip, bb= 0 0 1393 2477]{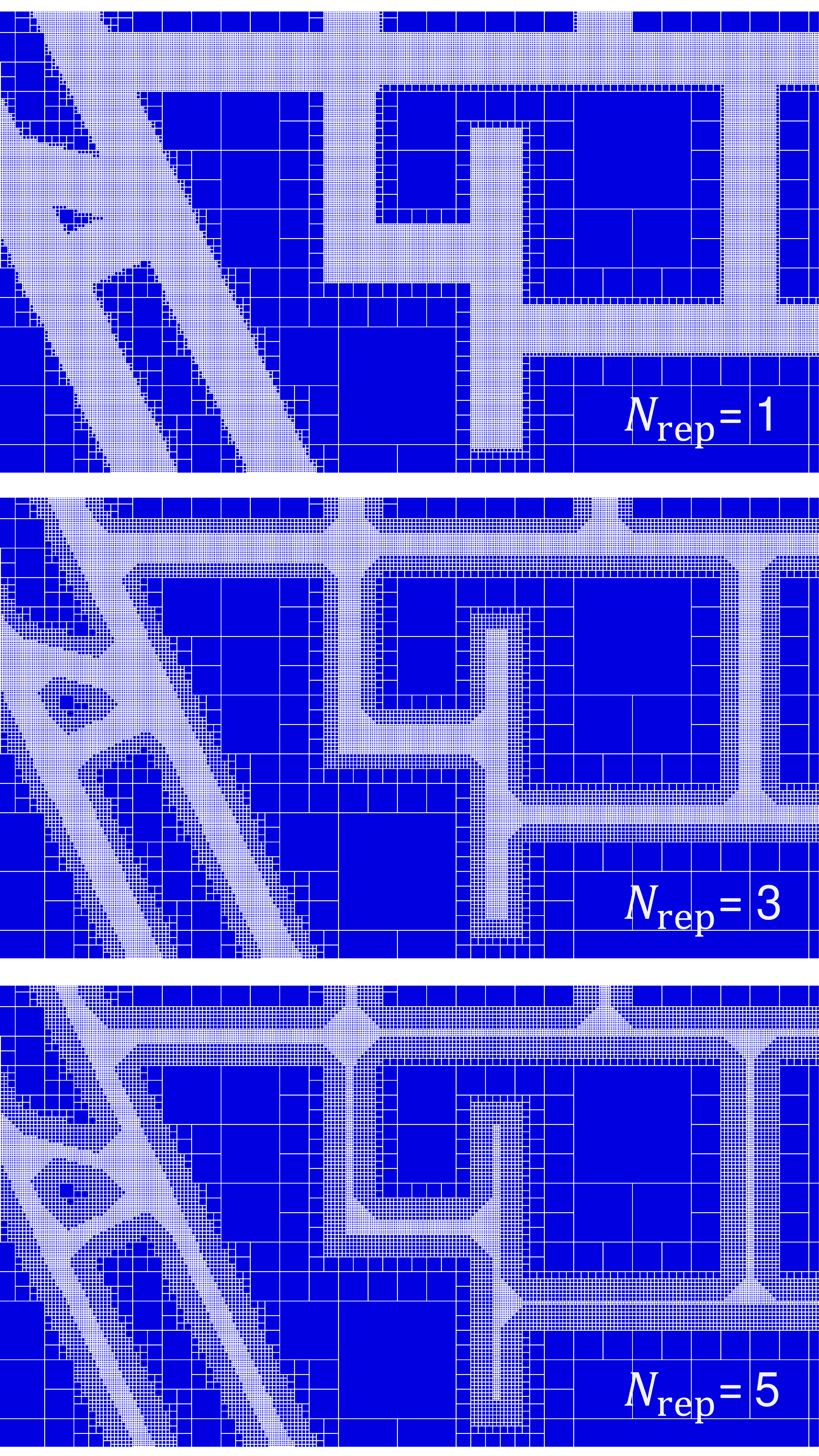}
\caption{Shrinking process of the areas around the centerline for different values of parameter $N_{\rm rep}$ between 1 and 5.}
\label{fig:shrinkingbynp}
\end{figure}

Let us define $N_{\rm rep}$ as the number of times the shrinking process is repeated and $N_{\rm rep}^{\rm max}$ as the number of times it is repeated until $F_{\rm stop}$ changes from false to true. Practically, repeating the shrinking process $N_{\rm rep}^{\rm max}$ times often causes over-release of the leaves. In the example shown in Fig.~\ref{fig:shrinkingbynp}, the parameter $N_{\rm rep}^{\rm max}$ becomes eighteen while it is adequate to set $N_{\rm rep}$ to six for generating a centerline graph. Hence, it is reasonable to regard the parameter $N_{\rm rep}^{\rm max}$ as a limit of iteration; at present, the actual number of $N_{\rm rep}$ needs to be adjusted for individual cases in the range from zero to $N_{\rm rep}^{\rm max}$.

\begin{figure}[t]
\vspace{-3.5cm}
\hspace{+2.5cm}
\includegraphics[width=0.85\textwidth, clip, bb= 0 0 1858 1393]{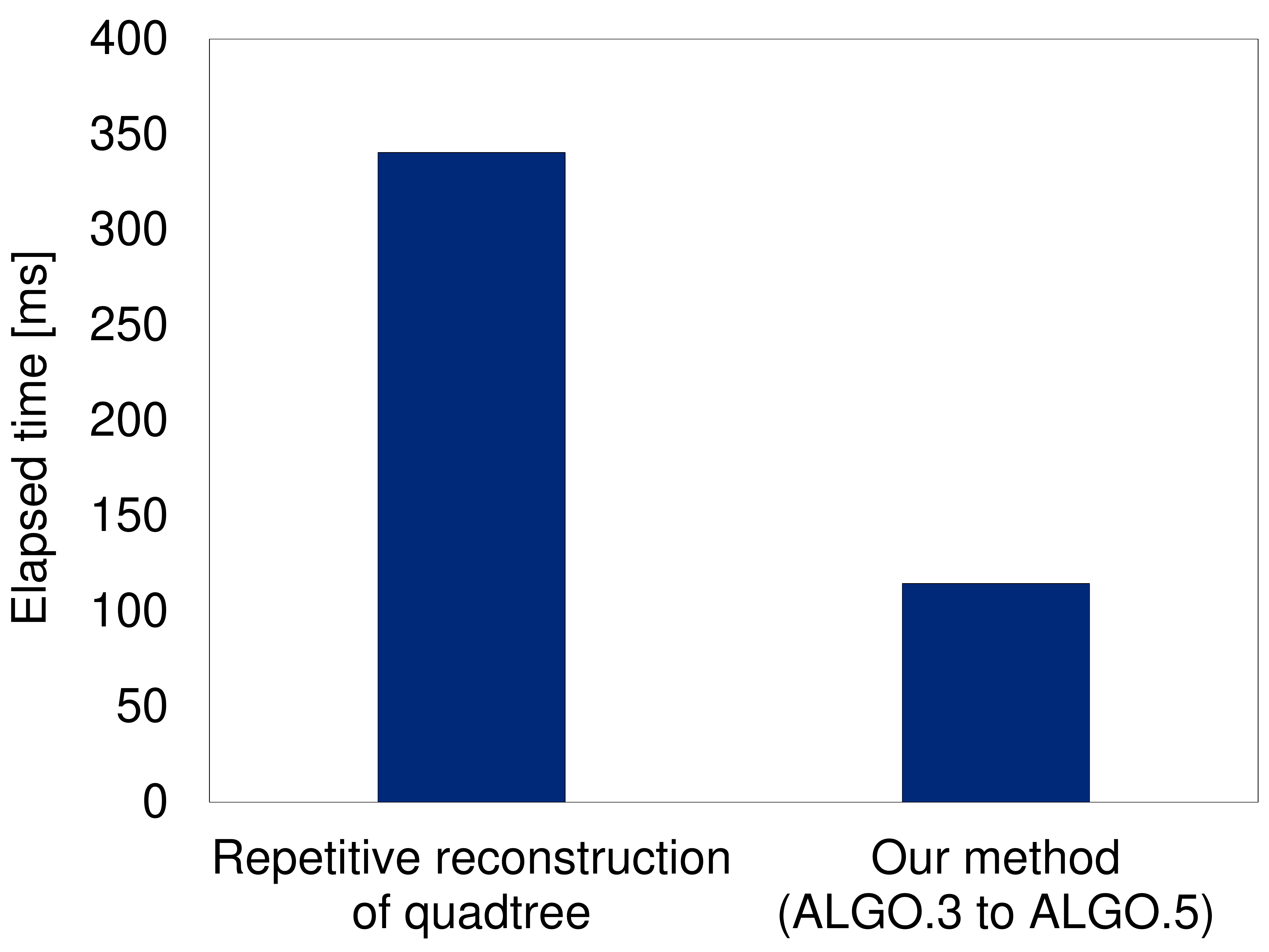}
\caption{Comparison of the computational time between the cases of repetitive tree reconstruction (left part) and our method (right part).}
\label{fig:comparetime}
\end{figure}

\HL{The shrinking process by using Algorithms~\ref{pesuedo-pointertable} to \ref{pesuedo-release} requires less computational time compared to the case of reconstructing quadtrees in every step of the $N_{\rm rep}$ iterations. Figure~\ref{fig:comparetime} shows a comparison of the computational time} \HL{consumed for the shrinking process between the repetitive reconstruction of quadtree and the proposed method, on a single core of an Intel Core i7-8650U CPU (1.90GHz). The result shows that our approach reduces the computational time by about 2.95 times compared to the repetitive tree reconstruction-based method.} 

\subsection{Selection and connection of nodes}\label{sec:selectconnectnodes}
After shrinking the road areas using Algorithm~\ref{pesuedo-release}, we trace all the remaining distal leaves of the tree by Morton's curve to choose the nodes from these distal leaves, which keep a certain distance, from among the previously selected nodes as described in Algorithm~\ref{pesuedo-selectnodes}. To begin with, we detect one of the distal leaves by using the procedure in lines2-6. After finding a distal leaf, we check the distance between the leaf and all the already registered nodes in lines8-14.  Here, the parameter $N_{\rm chk}^{\rm crr}$ represents the number of registered nodes. If all the distances between the leaf and each node registered at that time become larger than a certain distance $\eta$, we change the $F_{\rm chk}$ of the leaf from false to true and add the leaf to the list of registered nodes. After that, we increment the parameter $N_{\rm chk}^{\rm crr}$ by one as shown in line18.

\begin{figure}[t]
\vspace{-1.5cm}
\hspace{+3.5cm}
\includegraphics[width=0.8\textwidth, clip, bb= 0 0 1280 1280]{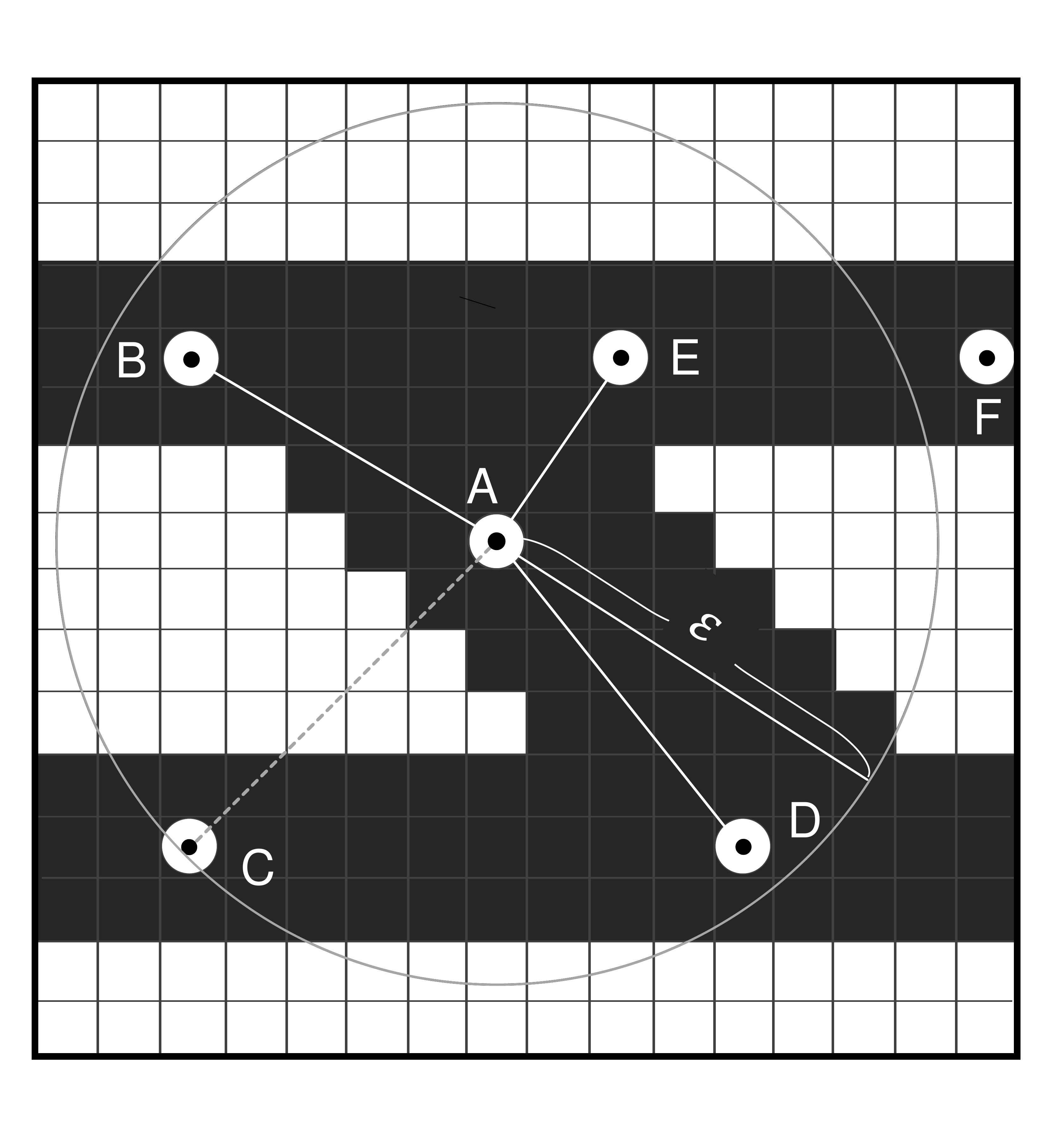}
\caption{Schematic view of connecting nodes.}
\label{fig:schemconnectnodes}
\end{figure}

Algorithm~\ref{pesuedo-connectnodes} describes the procedure of connection of nodes, and Figure~\ref{fig:schemconnectnodes} shows its schematic view. First, we define a cut-off radius $\epsilon$. Each node searches the other nodes located within the range $\epsilon$ as in lines5-6 of Algorithm~7. When the node $j$ finds that another node $i$ satisfies this condition, the node $j$ checks whether the relative vector from node $j$ to node $i$ only traverses the cells on the roads as shown in lines7-18. If it is true, the node $j$ adds the node $i$ to its connectivity list as shown in lines19-21.

The cut-off radius $\epsilon$ is the parameter that controls the degree of connectivity among the nodes; we experimentally determine it within a few times as large as the parameter $\eta$ in typical cases to strengthen the connectivity at intersections while keeping the geometry of the original roadmap. To summarize, we have two deterministic parameters and two experimental parameters: the number of cells in one direction $N_{\rm r}$, the distance among nodes $\eta$, the number of times of repeating the shrinking process $N_{\rm rep}$, and the cut-off radius $\epsilon$.

\begin{figure}[t]
\vspace{-3.5cm}
\hspace{+0.0cm}
\includegraphics[width=1.35\textwidth, clip, bb= 0 0 2477 1393]{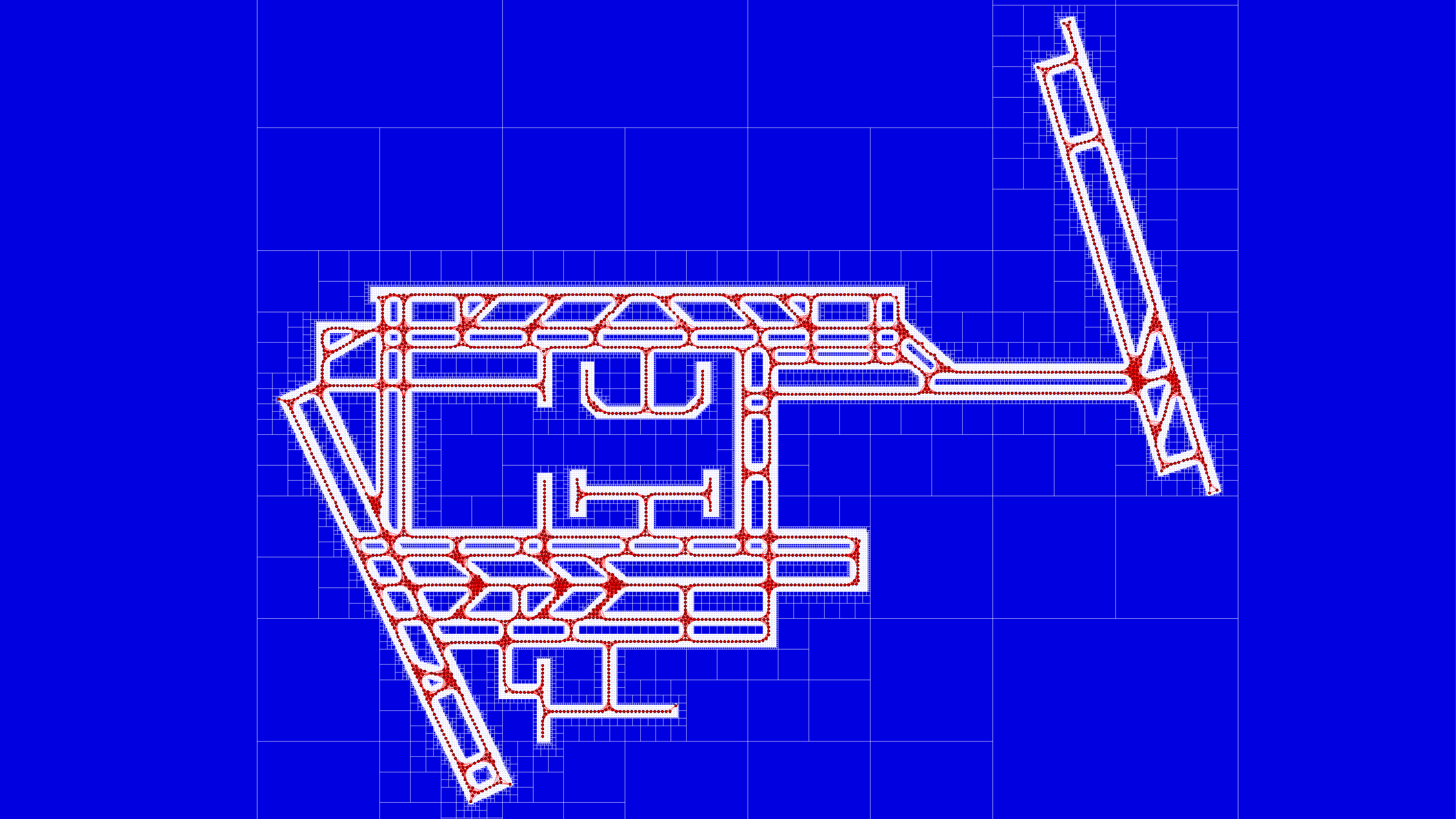}
\caption{A generated graph from the roadmap of Fig.~\ref{fig:schemclgraph} by using the procedures in Algorithms 1 to 7.}
\label{fig:GenGraphFromMap}
\end{figure}

\begin{figure}[t]
\vspace{-3.5cm}
\hspace{+0.0cm}
\includegraphics[width=1.35\textwidth, clip, bb= 0 0 2477 1393]{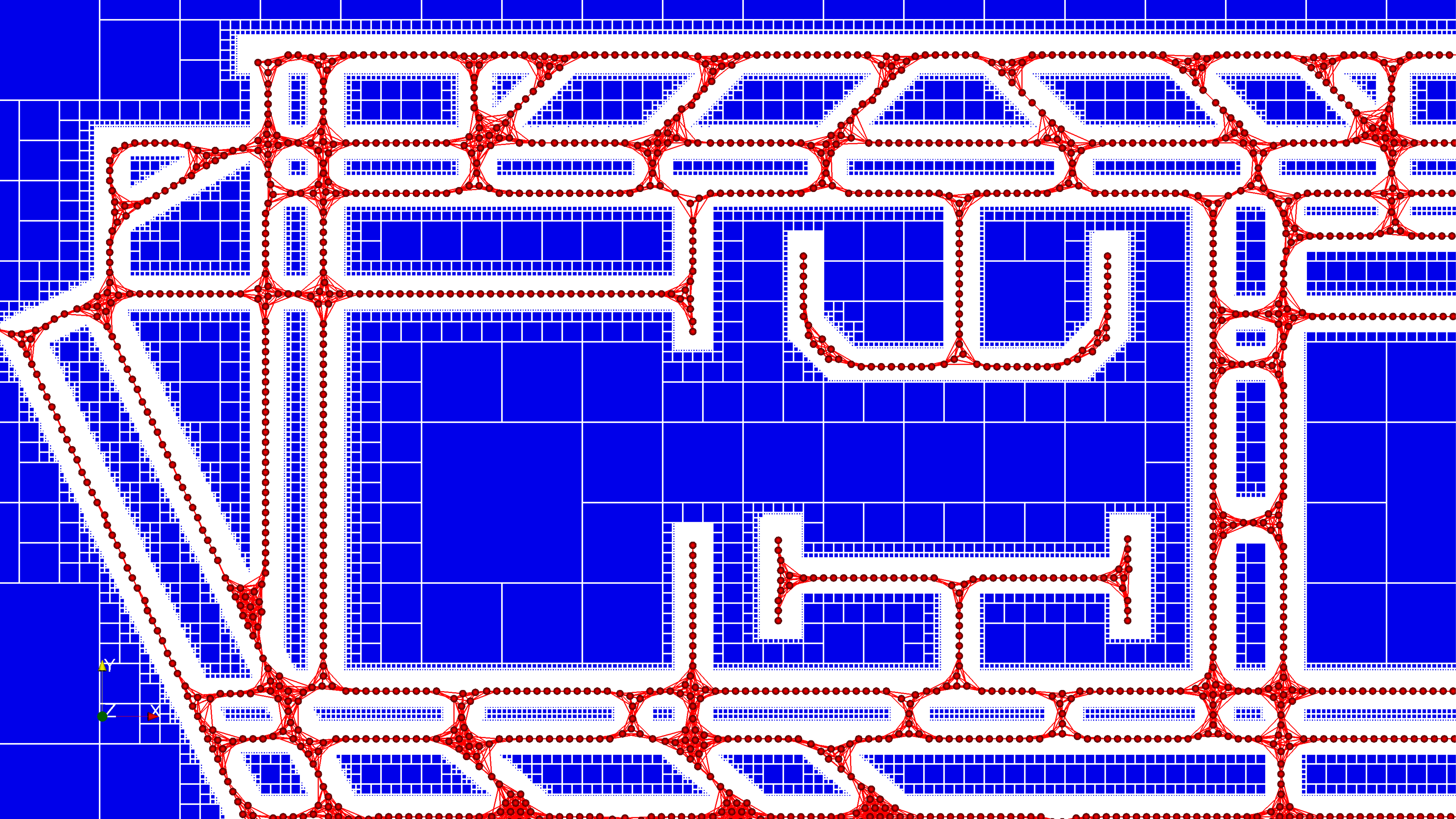}
\caption{An enlarged view of Fig.\ref{fig:GenGraphFromMap}.}
\label{fig:MeshandGraph}
\end{figure}

Figure~\ref{fig:GenGraphFromMap} shows a graph generated from the roadmap of Fig.~\ref{fig:schemclgraph} by using the procedures in Algorithms 1 to 7, and Figure~\ref{fig:MeshandGraph} shows an enlarged view of Fig.~\ref{fig:GenGraphFromMap}. The white-colored rectangle lines indicate the hierarchically structured leaves at a depth of zero, for reference. We set the pair of $(N_{\rm r}, \eta, N_{\rm rep}, \epsilon)$ to be $(2048, 8, 8, 28)$ to generate the graph. It was confirmed that the paths were constructed near the centerlines for the straight roads and multiple lanes were connected around each intersection.

\section{GRAPH NETWORK ANALYSIS}
\subsection{A short path search using Dijkstra's algorithm}
\begin{figure}[t]
\vspace{-3.5cm}
\hspace{+0.0cm}
\includegraphics[width=1.35\textwidth, clip, bb= 0 0 2477 1393]{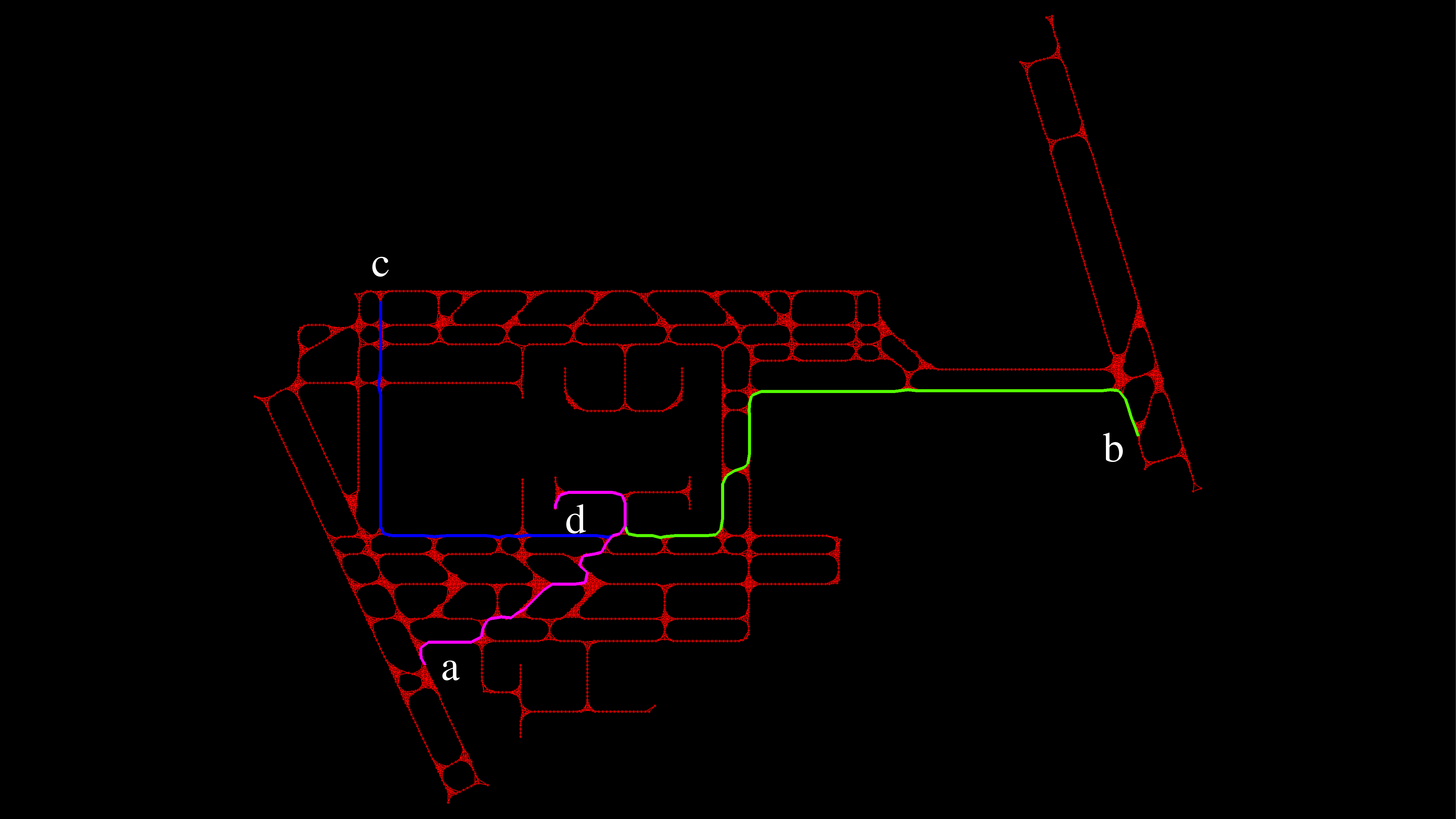}
\caption{The shortest paths from node $a$ to node $b$ or node $c$ via node $d$ calculated by using Dijkstra's algorithm.}
\label{fig:ShortPathDijkstra1}
\end{figure}

\begin{figure}[t]
\vspace{-3.5cm}
\hspace{+0.0cm}
\includegraphics[width=1.35\textwidth, clip, bb= 0 0 2477 1393]{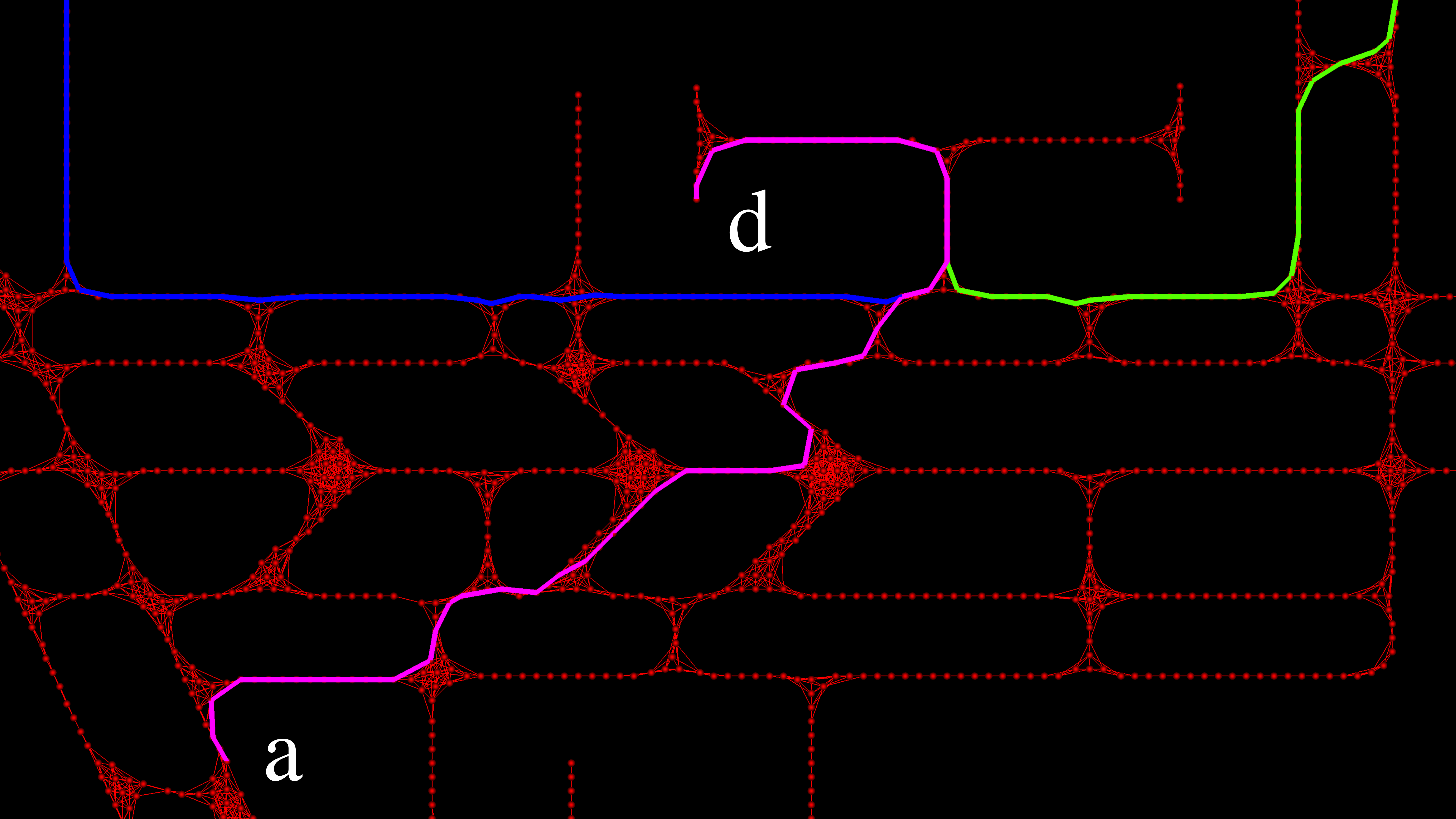}
\caption{An enlarged view of Fig.\ref{fig:ShortPathDijkstra1}.}
\label{fig:ShortPathDijkstra2}
\end{figure}

~\HL{As a case study}, we carry out a shortest-path finding using Dijkstra's algorithm for a network generated from the Tokyo International Airport in Japan. In one example, we calculate the shortest paths from the entrance of the high-speed taxiway in runway B to the departure lane of runway C or that of runway D via a checkpoint in Terminal~1. Figure~\ref{fig:ShortPathDijkstra1} shows the calculated paths from node $a$ to node $b$ or node $c$ via node $d$ by using Dijkstra's algorithm, and Figure~\ref{fig:ShortPathDijkstra2} shows an enlarged view of Figure~\ref{fig:ShortPathDijkstra1}. Several reasonable results are observed. First, the calculated paths show a characteristic of Euclidean distance rather than Manhattan distance because of the high-resolution image of the roadmap. We confirmed that the path connecting node $a$ and node $d$ crosses the runway A as diagonally as possible. This is understandable because of the ``triangle inequality'' relation among the three edges of a rectangle, the diagonal vertices of which are given as node $a$ and node $b$. A similar explanation can be made for the fact that the path connecting node $d$ and node $c$ shows two straight lanes including a single corner.\\

\begin{algorithm}[t]
\caption{Selection of Candidates}\label{pesuedo-cand}
\begin{algorithmic}[1]
\Function{selectCandidate}{$\rm Leaf$}
\If{${\rm Leaf.}F_{\rm end}$~$=$~${\rm false}$} 
\For{k~=~0,1...$N_{\rm child}-1$}
\State \Call{selectCandidate}{$\rm Leaf.child[k]$}
\EndFor
\Else
\If{$\rm Leaf.D$~$=$~$D_{\rm max}^{\rm mes}$}
\State flag $\leftarrow$ false
\State i $\leftarrow$ ${\rm Leaf.}L_{\rm min}.x$ 
\State j $\leftarrow$ ${\rm Leaf.}L_{\rm min}.y$
\State d $\leftarrow$ (Table[i+1][j].D)
\If{$\rm d$~$<$~$D_{\rm max}^{\rm mes}$}
\State flag $\leftarrow$ true
\EndIf
\State d $\leftarrow$ (Table[i-1][j].D)
\If{$\rm d$~$<$~$D_{\rm max}^{\rm mes}$}
\State flag $\leftarrow$ true
\EndIf
\State d $\leftarrow$ (Table[i][j+1].D)
\If{$\rm d$~$<$~$D_{\rm max}^{\rm mes}$}
\State flag $\leftarrow$ true
\EndIf
\State d $\leftarrow$ (Table[i][j-1].D)
\If{$\rm d$~$<$~$D_{\rm max}^{\rm mes}$}
\State flag $\leftarrow$ true
\EndIf
\If{flag~$=$~true}
\State ${\rm Leaf.}F_{\rm del}$~$\leftarrow$~${\rm true}$
\EndIf
\EndIf
\EndIf
\EndFunction
\end{algorithmic}
\end{algorithm}
\begin{algorithm}[t]
\caption{Sum of $F_{\rm del}$ over a Bundle of Sub-leafs}\label{pesuedo-sumbundleleafs}
\begin{algorithmic}[1]
\Function{checkBranches}{$\rm Leaf$}
\If{$\rm Leaf.D$~$\ne$~$D_{\rm max}^{\rm mes}$ and \\~~~~~~~~~~$\rm Leaf.D$~$\ne$~$D_{\rm max}^{\rm mes}-1$}
\For{k~=~0,1...$N_{\rm child}-1$}
\State \Call{checkBranches}{$\rm Leaf.child[k]$}
\EndFor
\Else
\If{${\rm Leaf.D} = D_{\rm max}^{\rm mes}-1$}
\State ${\rm Leaf.}S_{\rm del}$~$\leftarrow$~${\rm 0}$
\For{k~=~0,1...$N_{\rm child}-1$}
\If{$\rm Leaf.child$[k].$F_{\rm del}$ = true}
\State ${\rm Leaf.}S_{\rm del}$~$\leftarrow$~${\rm Leaf.}S_{\rm del}$+1 
\EndIf
\EndFor
\EndIf
\EndIf
\EndFunction
\end{algorithmic}
\end{algorithm}

\subsection{Discussions for future work}
\begin{figure}[t]
\vspace{-7.5cm}
\includegraphics[width=1.3\textwidth, clip, bb= 0 0 960 1280]{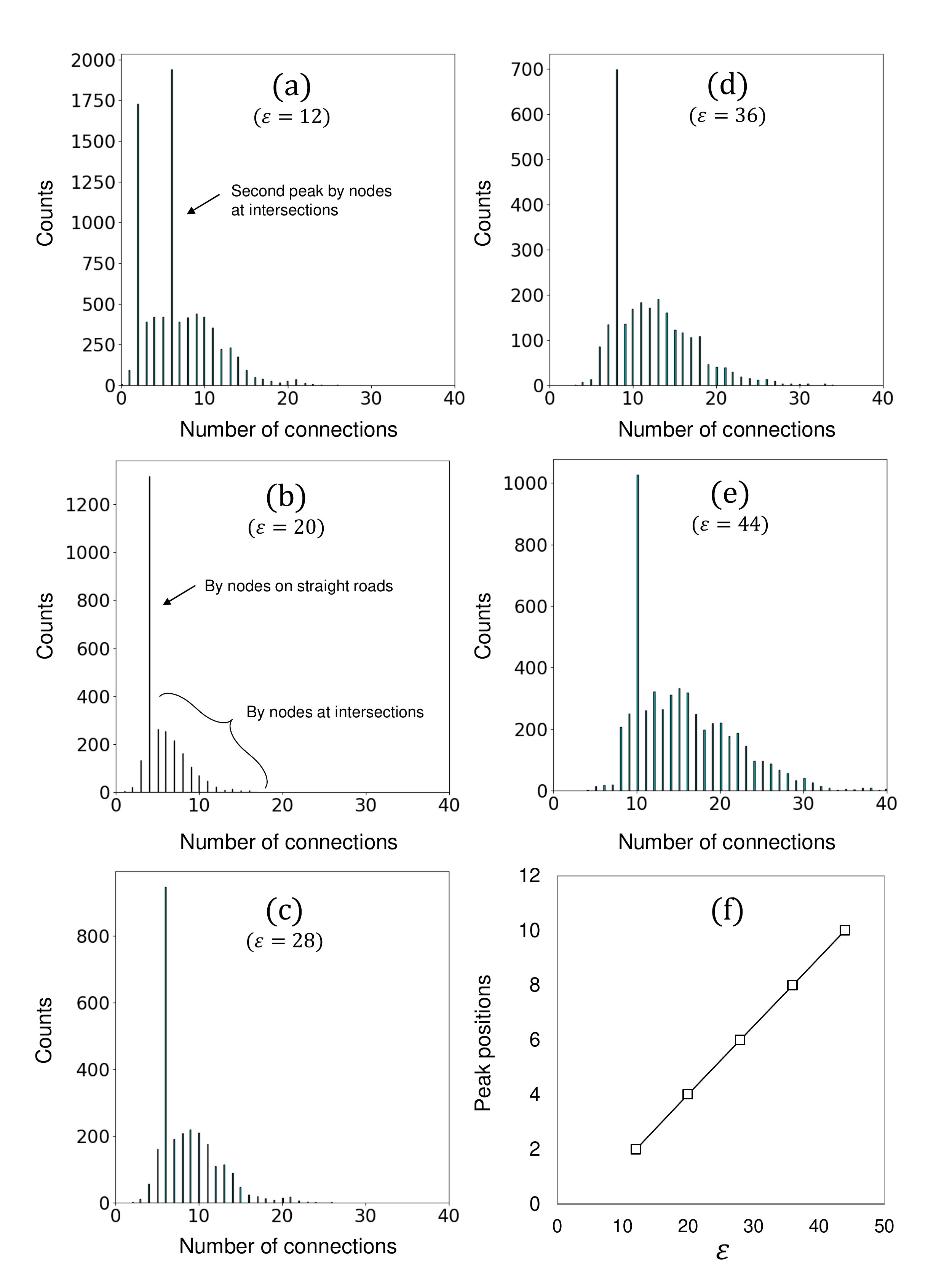}
\caption{The histograms of node-connectivity.}
\label{fig:histnodeconnect}
\end{figure}

~\HL{As mentioned in the Introduction, this paper aims to conduct a preliminary study as the first step toward the auto-generation of the network graph used in practical traffic CA simulations. To discuss the applicability of our method to traffic CA simulations quantitatively, we examine the node connectivity of the generated graph in Fig.\ref{fig:ShortPathDijkstra2}. Figures~\ref{fig:histnodeconnect}(a) to (e) show the histograms of the node connections for different $\epsilon$ between 12 and 44; the horizontal axis indicates the number of connections per node, and the vertical axis shows their distributions in each case. In Fig.\ref{fig:histnodeconnect}(c), the nodes on the straight roads contribute to the emergence of a peak at the value of six, which indicates that the nodes on the straight lines connect each other not only between the neighboring nodes but also among three nodes away in the front} \HL{and back directions. It can be said that this redundant connectivity could enrich the functionality of the applications such as bypass roads or passing phenomena. Furthermore, Fig.\ref{fig:histnodeconnect}(f) shows the dependence of the positions of the peaks generated by the nodes of the straight roads on the parameter $\epsilon$. Figure~\ref{fig:histnodeconnect} clearly shows that the number of connections between nodes on straight roads is proportional to the parameter $\epsilon$.

Meanwhile, the nodes at the intersections contribute to the moderate distribution at the lower part of each case. The dispersion of each distribution directly correlates to that of the connections at each node; thus, the sharp distribution provides a high uniformity for the number of selections of paths at intersections. From Fig.\ref{fig:histnodeconnect}(a), it can be seen that most of the nodes at the intersections connect with two nodes because of the small $\epsilon$, and therefore the second peak emerges (here, we several nodes disconnect from each other because the parameter $\epsilon$ is too small; as for the setting of the parameter $\epsilon$, we must adjust the parameter $\epsilon$ experimentally as described in Section~\ref{sec:selectconnectnodes}).
In any case, the histograms in Fig.\ref{fig:histnodeconnect} give an objective view of the generated graphs. Further comparisons of these histograms in different related methods must be conducted in a future work.}

\section{CONCLUSIONS}
In this paper, we proposed an effective method for the auto-generation of a centerline graph from geometrically complex roadmaps of real-world traffic systems by using a hierarchically structured tree.

Our method is summarized as follows. First, we store the binary values of a solid image of a roadmap in a two-dimensional square map. Second, we recursively divide the square map into sub-leafs by a quadtree until the summed-up value of pixels included inside the leaf becomes equal to or less than one. Third, we gradually remove the distal leaves that are adjacent to the leaves whose depths are shallower than the distal leaf during the shrinking process. After that, we trace the remaining distal leaves of the tree using Morton's curve, while selecting the leaves that keep a certain distance from among the previously selected leaves as the nodes of the graph. In the end, each selected node searches the neighboring nodes and stores them as the edges of the graph.

The features of our method are as follows: (1) a straightforward method using only a hierarchical tree without combining any other supportive algorithms such as distance function, (2) the direct generation from a solid image, and (3) multiple lanes at each intersection. 
Besides, our method significantly improves computational efficiency compared to the existing algorithms; it requires only a one-time recursive tree construction. This is a significant point compared to the previous studies. 

Additionally, we demonstrated the shortest-path findings of the generated graph using Dijkstra's algorithm. It is quite meaningful that we successfully established a methodology to generate a network graph that is suitable for the studies of vehicular traffic CA simulations.

\section*{Acknowledgements}
This research was supported by MEXT as ``Post-K Computer Exploratory Challenges'' (Exploratory Challenge 2: Construction of Models for Interaction Among Multiple Socioeconomic Phenomena, Model Development and its Applications for Enabling Robust and Optimized Social Transportation Systems)(Project ID: hp190163), partly supported by JSPS KAKENHI Grant Numbers 25287026, 15K17583 and 18H06459.
We would like to thank Editage (www.editage.jp) for English language editing.
\\\\
\bibliographystyle{h-physrev3}
\bibliography{reference}

\begin{algorithm}[b]
\caption{Release of leaves}\label{pesuedo-release}
\begin{algorithmic}[1]
\Function{releaseLeaves}{$\rm Leaf$, $F_{\rm stop}$}
\If{$\rm Leaf.D$~$\ne$~$D_{\rm max}^{\rm mes}$ and \\~~~~~~~~~~$\rm Leaf.D$~$\ne$~$D_{\rm max}^{\rm mes}-1$}
\For{k~=~0,1...$N_{\rm child}-1$}
\State \Call{releaseLeaves}{$\rm Leaf.child[k]$,$F_{\rm stop}$}
\EndFor
\Else
\If{${\rm Leaf.D} = D_{\rm max}^{\rm mes}-1$}
\State $F_{\rm nbr}$~$\leftarrow$~${\rm false}$
\For{k~=~0,1...$N_{\rm child}-1$}
\State i $\leftarrow$ ${\rm Leaf.child[k].}L_{\rm min}.x$ 
\State j $\leftarrow$ ${\rm Leaf.child[k].}L_{\rm min}.y$
\ForAll{$\rm J$ s.t. ${\rm j-1}\leq {\rm J}\leq {\rm j+1}$}
\ForAll{$\rm I$ s.t. ${\rm i-1}\leq {\rm I}\leq {\rm i+1}$}
\If{(i,j)$\ne$(I,J)}
\State d $\leftarrow$ Table[I][J].D
\If{ d = $D_{\rm max}^{\rm mes}$}
\State s $\leftarrow$ Table[I][J].parent.$S_{\rm del}$ 
\If{s = 0}
\State $F_{\rm nbr}$~$\leftarrow$~${\rm true}$
\EndIf
\EndIf
\EndIf
\EndFor
\EndFor
\EndFor
\State $F_{\rm cnd}$~$\leftarrow$~${\rm false}$
\For{k~=~0,1...$N_{\rm child}-1$}
\If{${\rm Leaf.child[k].}F_{\rm del}$ = true}
\State $F_{\rm cnd}$~$\leftarrow$~${\rm true}$
\EndIf
\EndFor
\If{$F_{\rm nbr}$ = true and $F_{\rm cnd}$ = true}
\State Leaf.$F_{\rm end}$ $\leftarrow$ true
\For{k~=~0,1...$N_{\rm child}-1$}
\State delete Leaf.child[k]
\State $F_{\rm stop}$ $\leftarrow$ true
\EndFor
\EndIf
\EndIf
\EndIf
\EndFunction
\end{algorithmic}
\end{algorithm}

\begin{algorithm}[t]
\caption{Selection of Nodes}\label{pesuedo-selectnodes}
\begin{algorithmic}[1]
\Function{selectNodes}{$\rm Leaf$}
\If{${\rm Leaf.}F_{\rm end}$~$=$~{\rm false}}
\For{k~=~0,1...$N_{\rm child}-1$}
\State \Call{selectNodes}{$\rm Leaf.child[k]$}
\EndFor
\Else
\If{${\rm Leaf.D} = D_{\rm max}^{\rm mes}$}
\State $F$ $\leftarrow$ true
\For{j~=~0,1...$N_{\rm chk}^{\rm crr}-1$} 
\State d = \Call{distance}{$\rm node[j].pos$, $\rm Leaf.pos$} 
\If{d $<$ $\eta$}
\State $F$ $\leftarrow$ false
\EndIf
\EndFor
\If{$F$~$=$~true}
\State ${\rm Leaf.}F_{\rm chk}$ $\leftarrow$ true
\State add Leaf to the list of nodes
\State $N_{\rm chk}^{\rm crr}$ $\leftarrow$ $N_{\rm chk}^{\rm crr}+1$
\EndIf
\EndIf
\EndIf
\EndFunction
\end{algorithmic}
\end{algorithm}

\begin{algorithm}[t]
\caption{Connection of selected nodes}\label{pesuedo-connectnodes}
\begin{algorithmic}[1]
\For{j~=~0,1...$N_{\rm node}-1$}
\State $\rm node[j].list \leftarrow null$
\For{i~=~0,1...$N_{\rm node}-1$}
\If{$\rm i~\ne~j$}
\State $\rm d$ = \Call{distance}{$\rm node[j].pos$, $\rm node[i].pos$} 
\If{$\rm d < \epsilon$}
\State define $A$ as a relative vector from $\rm j$ to $\rm i$
\State define $n$ as a unit vector of $A$
\State k $\leftarrow$ 0
\State pos $\leftarrow$ ${\rm node[j].pos}$
\State flag $\leftarrow$ true
\While{${\rm pos} \ne {\rm node[i].pos}$}
\State pos $\leftarrow$ pos $+$ ${\rm k}n$
\If{$B_{\rm p}~{\rm at}~{\rm pos}~=~0$}
\State flag $\leftarrow$ false
\EndIf
\State k $\leftarrow$ k$+$1
\EndWhile
\If{${\rm flag} = {\rm true}$}
\State add $\rm node[i]$ to $\rm node[j].list$
\EndIf
\EndIf
\EndIf
\EndFor
\EndFor
\end{algorithmic}
\end{algorithm}

\clearpage
\clearpage



\end{document}